# Repulsive interactions of eco-corona covered microplastic particles quantitatively follow modelling of polymer brushes

**Thomas Witzmann**[1], Anja F. R. M. Ramsperger[2, 3], Simon Wieland[2, 3], Christian Laforsch[2], Holger Kress[3], Andreas Fery[1, 4], and Günter K. Auernhammer[1]

[1] Leibniz-Institute of Polymer Research Dresden, Hohe Straße 6, 01069 Dresden, Germany

[2] Animal Ecology I and BayCEER, University of Bayreuth, 95447 Bayreuth, Germany

[3] Biological Physics Group, University of Bayreuth, 95447 Bayreuth, Germany

[4] Physical Chemistry of Polymeric Materials, Technische Universität Dresden, Bergstr. 66, 01069 Dresden, Germany

Corresponding authors: fery@ipfdd.de, auernhammer@ipfdd.de

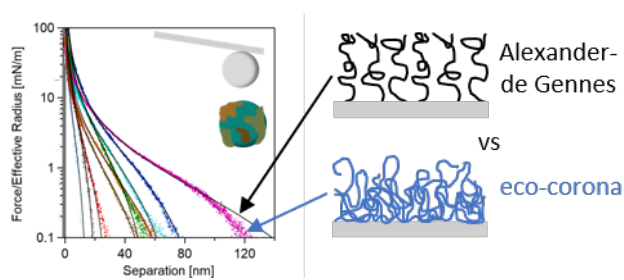

*For Table of Contents use only*

## Abstract

Environmental fate and toxicity of microplastic particles is dominated by their surface properties. In the environment an adsorbed layer of biomolecules and natural organic matter forms the so-called eco-corona. A quantitative description of how this eco-corona changes the particles' colloidal interactions is still missing. Here, we demonstrate with colloidal probe-atomic force microscopy that the formation of the eco-corona on microplastic particles introduces a soft film on the surface which changes the mechanical behaviour. We measure single particle-particle interactions and find a pronounced increase of long-range repulsive interactions upon eco-corona formation. These force-distance characteristics follow well the polymer brush model by Alexander and de Gennes. We further compare the obtained fitting parameters to known systems like polyelectrolyte multilayers and propose these as a model system for the eco-corona. The foundation of the eco-corona interacting like a polymer brush with its surrounding may help understand microplastic transport and aggregation in the environment.

## 1. Introduction

Plastic litter is an environmental concern as it is ubiquitous in the water, air and on land.[1-5] Exposed to environmental conditions like changing temperature, mechanical stress or UV radiation it degrades into smaller particles.[6, 7] Next to these particles, abrasion from any plastic product and particles generated in the size range below 5mm are termed as microplastics.[8] Its ubiquitous abundance and small size pose a risk to organisms and human health due to particle ingestion and inhalation.[9, 10] In the last decade toxicity tests on various organisms and cells with different polymer types, sizes and shapes were performed.[11-13] Most of the studies focused thereby on spherical model particles, which are either pristine (plain) or functionalised. Although it is necessary to start investigations with well-defined particles, results of such studies do not represent the properties of real environmental microplastics.[14]

As the particles disintegrate and shrink in size, their surface area to volume ratio increases. Therefore, the surface is dominating the interaction with its surrounding making it essential to consider its properties. In addition to the increased surface area the underlying mechanisms that cause disintegration do also change the surface properties.[7, 15] But not only disintegration lead to surface modification. In the moment microplastic particles enter the environment they are exposed to a mix of different biomolecules (proteins, polysaccharides) and natural organic matter (NOM) like humic acids. These molecules adsorb to the particles surface forming an eco-corona.[16] The adsorption of these molecules alters eventually the microplastics surface morphology, charge, chemistry and potentially mechanics.[17-20]

Several aspects are affected when the surface changes its properties. Singh et al.[21] investigated the aggregation behaviour of polystyrene (PS) nanoparticles. They demonstrated in the presence of humic acids that these particles are stabilised due to steric forces upon adsorption of these molecules. In accordance to these results Wu et al.[22] also found humic acids to stabilise PS nanoplastics in solution preventing aggregation.

Another aspect affected by eco-corona formation is particle-cell interaction. Ramsperger et al[23]. showed that the induced change of the surface properties leads to an increase of particle-cell attachment and uptake. The eco-corona and its effects on the cell interactions might therefore also change toxicity of microplastic particles. Whether surface charge, mechanics, morphology or chemistry or the contribution of several factors play the dominant role for changes in particle-cell interactions is yet unknown. The role of mechanical properties of particles on cellular uptake has received attention recently. Hartmann et al[24] investigated with the help of colloidal probe-atomic force microscopy (CP-AFM) the stiffness-dependent uptake of microparticles into cells. They could show that softer particles are transported faster to lysosomes than stiffer ones. Due to the adsorption of hydrated molecules eco-corona covered particles are expected to be softer at the surface than plain particles. Therefore, stiffness-dependent uptake is expected to be important for eco-corona covered microplastics as well. Examining the mechanical properties before and after the eco-corona formation is crucial. It clarifies the impact of environmental exposure onto the particle mechanics properties and opens the way to study how cellular uptake is affected.

Direct force measurements on microplastic particles covered by an eco-corona have not been performed, let alone measurements that compare particles prior and after eco-corona formation. We thus carried out a systematic study on particles. The microplastic particles have been exposed to a saltwater environment, since it is assumed as one potential sink of pollution.[25] Subsequently we measured with CP-AFM single particle-particle interaction forces from 10 pN to several 10 nN with a vertical resolution of 20 pm and a lateral resolution of 50 nm. This means that attractive or repulsive forces occurring between colloidal probe and microplastic particles can be detected. Interaction forces are measurable during approach and retraction of the probe from the sample particle. To measure only mechanical repulsive interactions with CP-AFM but in a realistic environment we focused on physiological salt concentrations. With this we reveal the changes induced by the eco-corona formation and describe quantitatively the repulsive interactions using established models for polymer (brush) surface layers. This work helps to understand and describe the physical structure of the eco-corona. It therefore gives a quantitative measure to improve and implement well-controlled model systems for environmental microplastic particles.

## 2. Methods and materials

### 2.1. Microplastic particles and pre-treatment conditions

The particles were incubated as described in Ramsperger et al..[23] In brief, 3 µm sized, spherical, non-functionalised polystyrene microplastic particles (Micromod, Rostock, Germany, white particles, micromer plain, Prod. Nr. 01-00-303) were incubated in saltwater to allow the formation of an eco-corona. Therefore, 20 µL of the polystyrene stock solution was given into a glass vial and 980 µL salt water was added. The salt water was taken from a highly biodiverse coral reef aquarium with a salinity of 3.5 % and was exchanged by new media three times per week via centrifugation (2000g, 20 min at room temperature). The procedure was repeated for two weeks and afterwards the microplastic particles were investigated with scanning electron microscopy (SEM) and CP-AFM.

### 2.2. Colloidal probe-atomic force microscopy (CP-AFM)

Direct force measurements were conducted with a MFP-3D Bio (Asylum Research Inc., Santa Barbara, USA) mounted on an inverted optical microscope (Axio Observer Z1, Zeiss, Oberkochen, Germany). Before the measurements, tipless cantilevers (CSC38, MikroMasch, Sofia, Bulgaria) were calibrated according to the thermal noise method.[26, 27] Cantilevers had a spring constant of 0.24 N/m and 0.28 N/m. To prepare colloidal probes, cantilevers were rinsed in Milli-Q water, ethanol, Milli-Q water, and acetone and treated with air-plasma for 10 min (SmartPlasma, plasma technology GmbH, Herrenberg-Gülstein, Germany) before silica colloidal particles (nominal diameter of 1.76 µm, microParticles GmbH, Berlin, Germany) were attached to the cantilevers with 2-components epoxy glue (UHU Plus Endfest, UHU GmbH & Co. KG, Bühl/Baden, Germany).

Microplastic particles were allowed to sedimented onto glass slides in fluid cells for 60 minutes. For pristine microplastic particles, a precleaned glass slide was modified with 100 µL of 5 cSt silicone oil (Gelest Inc, Morrisville, USA) on a heating plate at 200 °C for 3 min. Afterwards it was sonicated in an acetone and ethanol bath for 10 min each to remove residual silicone oil. This surface modification is a modified version of Eifert et al.[28] and allowed fixation of the pristine microplastic particles to the glass substrate. Eco-corona particles are able to stick on the substrate without modification. The liquid cell was eventually rinsed three times with 150 mM aqueous KCl solution. Subsequently, CP-AFM measurements were performed with a tip velocity of 300 - 400 nm/s at a scan rate of 0.5 Hz in 150 mM aqueous KCl solution of pH 6-7. Force curves of 10 – 20 individual particles were evaluated for each sample.

With CP-AFM it is possible to measure interaction forces from about 10 pN to several N with a vertical resolution of 20 pm and a lateral resolution of 50 nm. CP-AFM works as following (see Figure 1) with the help of a colloidal probe, in this case a silica sphere, glued to a cantilever interaction forces are measured. Any attractive or repulsive force between the colloidal probe and the sample deflects the cantilever. The cantilever deflection is measured through a laser beam focused on the back of the cantilever and reflected on a quadrant photodiode for a sensitive position detection of the incident laser light.

In force vs. separation curves, positive force values indicate a repulsive force between the colloidal probe and the sample and negative forces values correspond to an attraction. Zero separation is defined at the incompressible contact of two hard materials, often called the constant compliance regime. Forces are detected when the colloidal probe is approaching the sample (approach curves) or when the colloidal probe is retracted from the sample (retraction curve). Approach curves give information about non-contact interaction like Van-der-Waals, electrostatic forces, and forces due to mechanical deformation or steric repulsion. The retraction curves describe additionally the adhesion

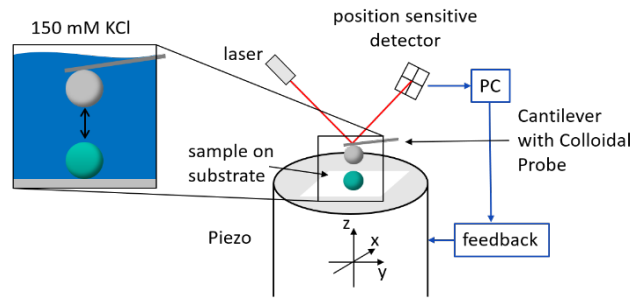

*Figure 1: Schematic setup of a colloidal probe-atomic force microscope. The interaction of a silica colloidal probe and a microplastic particle lead to deflection of the cantilever and are position sensitive detected by a laser. Attractive and repulsive contact and non-contact forces are detectable.*

between colloidal probe and sample.

The salt water which was used to incubate the eco-corona particles has a salinity of 35 ‰. This translates to a hypothetic KCl ion concentration of 469 mM. Comparing to the measurement solution of 150 mM KCl it is more than three times higher. The higher ion concentration in the natural system leads to a higher osmotic pressure. Therefore, smaller brush thicknesses are expected in the natural environment compared to the experimental results. We note, however, that the observed scaling and mechanisms are independent of these details. For both salt concentrations, electrostatic interactions can safely be neglected.

## 2.3. Scanning electron microscopy (SEM)

To investigate the eco-corona on the microplastic particles, we transferred 100 µl of the environmentally exposed microplastic particles on a coverslip (Ø 12mm, Menzel Gläser, Thermo Scientific). The eco-corona was then fixed using Karnovsky's fixative (2% PFA and 2.5% glutaraldehyde in 1x PBS) prior to dehydration in an ethanol series (30%, 50%, 70%, 80%, 90% for 30 min each, 95% and absolute ethanol for 1 h each). Then, the samples were dried in hexamethyldisilazane (HMDS, Carl Roth GmbH). The dry samples were then placed on carbon conductive tabs (Ø 12 mm, Plano GmbH, Wetzlar, Germany) fixed to aluminium stubs (Ø 12 mm, Plano GmbH, Wetzlar, Germany). Subsequently, the samples were coated with a 4 nm thick platinum layer (208HR sputter coater,

Cressington, Watford, UK). They were analysed with a scanning electron microscope (FEI Apreo VolumeScope) at 5 kV using an Everhart-Thornley detector.

## 3. Microplastic particle properties

### 3.1. Scanning electron microscopy (SEM)

The SEM images in Figure 2 show a clear difference between the pristine particles (Figure 2A) and particles that were incubated in salt water (Figure 2B). The pristine particles are monodisperse with a homogenous and smooth surface. The incubation of these particles in salt water for two weeks leads to the formation of an eco-corona. This eco-corona is heterogeneously distributed over the surface of the particles. While some particles are heavily covered others only show slight coverage.

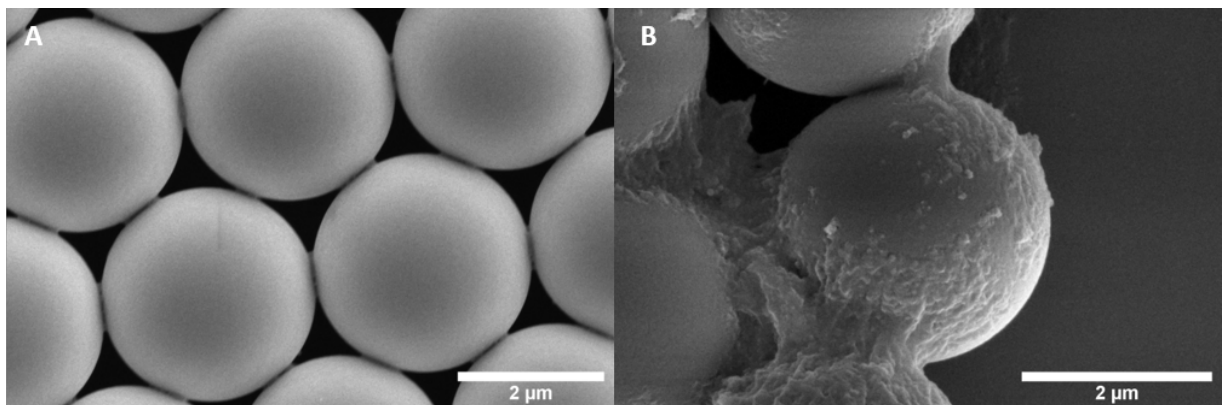

*Figure 2: Comparison of the surface morphology of pristine polystyrene microplastic particles (A) and the same type of particles bearing an eco-corona after incubation in salt water (B). Microplastic particles have a diameter of 3 µm.*

### 3.2. Direct force measurements with colloidal probe

Direct force measurements were conducted in an asymmetric fashion meaning between silica the colloidal probe and the microplastic particles. The experiment was performed in a 150 mM aqueous KCl solution. The high ionic strength induces an osmotic pressure that acts against the ions forming the electric double layer around a charged surface. This suppresses the electric double layer to such an extent that the theoretical electrostatic interaction range commonly known as Debye length is below 1 nm. Therefore, contributions of attractive or repulsive double layer forces can be neglected and only steric forces are present.

The CP-AFM approach curves on pristine and incubated particles differ substantially, as shown in Figure 3 (pristine microplastic particles 3A and eco-corona covered particles 3B). Force curves in 3A show typical mechanical interaction of two hard materials, in this case the pristine polystyrene particle and the silica colloidal probe. The repulsive force increases suddenly at low separation. After deforming the pristine particles asperities (RMS roughness = 1.27 nm), force curves reach the constant compliance regime, i.e., no further deformation is done. In contrast the data for eco-corona covered particles in graph 3B looks very different. Even at separations above 50 nm, there is already a small repulsive force detectable. This repulsion increases slowly during the further approach and will be discussed in detail later on. Such a behaviour is typical for compressing a soft material and is therefore called soft repulsion. In our case, the compression of the eco-corona is visible by the soft repulsion regime. Only below 10 nm the repulsive force increases more rapidly and eventually ends in the constant compliance regime of the hard contact at 0 nm. In this process, the eco-corona gets more and more compressed and finally the hard material of the underlying polystyrene is dominating the interaction. No time-dependent trend in the interaction behaviour, like a restructuring of the eco-corona, where observed over the experimental duration of 5 h.

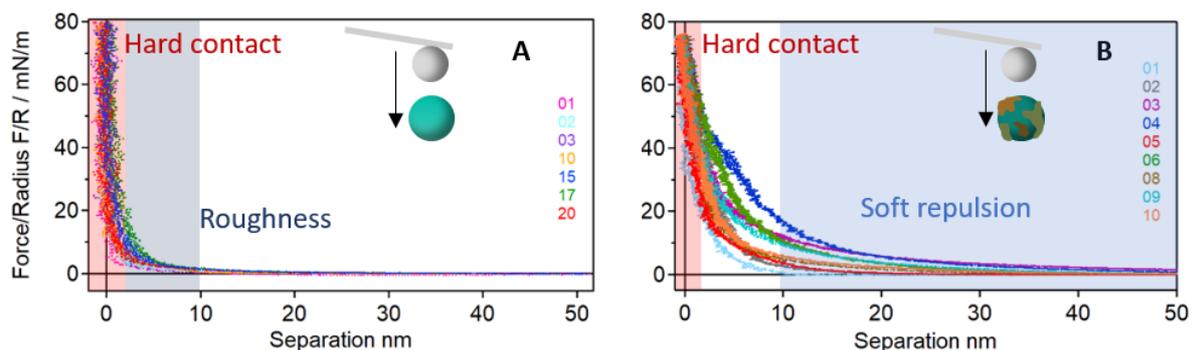

*Figure 3: Normalized force vs separation curves between the silica probe and pristine polystyrene particles (A) and between the silica probe and particles covered with an eco-corona (B) at 150 mM KCl. Approach curves show soft mechanical repulsion of the eco-corona in (B). For pristine particles in (A) no additional interaction forces were measured other than the one of the hard contact and roughness (RMS = 1.27 nm). Coloured numbers indicate the order in which particles were measured.*

In comparison with Figure 3A, the force curves in Figure 3B show more variations between the different particles. The range and magnitude of the soft repulsion are different for different particle. This can be explained by the general heterogeneous appearance of the eco-corona covered particles, also shown in the SEM images (Figure 2: Comparison of the surface morphology of pristine polystyrene microplastic particles (A) and the same type of particles bearing an eco-corona after incubation in salt water (B). Microplastic particles have a diameter of 3 µm.Figure 2). However, the force curves can give clear information whether a particle is covered with an eco-corona.

We find no systematic trend of the repulsive interactions dominating the approach data for repeated measurements on the same particles. This indicates elastic behaviour of the eco-corona. But we find a decreasing attractive force (**Error! Reference source not found.**) during approach of consecutive measured particles in the beginning of the experiment. This agrees with the data acquired on the retraction of the cantilever which shows an adhesion signature (**Error! Reference source not found.**) which decreases with the measurement on consecutive particles. This can be explained by transfer of eco-corona material to the probe particle which alters the particle-particle adhesion (see **Error! Reference source not found.**), but has only a negligible influence on the nanomechanical properties of the particle-particle contact.

## 4. Repulsive interactions of eco-corona particles: Quantitative description

The eco-corona is probably a highly swollen biomolecular material that is physisorbed to the particle surface. In a similar system Block and Helm observed that polyelectrolytes adsorbed from solutions with high ionic strength can be described by the Alexander-de Gennes (AdG) polymer brush theory.[29-32] This anchoring to a surface is an important similarity to polymer brushes. In contrast, the molecular structure is probably not brush-like, but rather unordered with some chemical or physical crosslinks in the material in the case of the eco-corona. Accordingly, we compare the measured force curves for the eco-corona to the prediction of the AdG model in Figure 4.

*Figure 4: Normalized force vs separation curves between silica probe and various eco-corona particles at 150 mM KCl. Black lines correspond to asymmetric Alexander-de Gennes fit at 293 K. Coloured numbers indicate in which order particles where measured.*

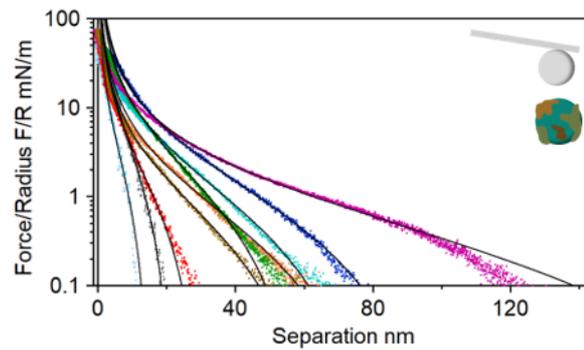

The model by Alexander and de Gennes[33, 34] can deal with polymer brushes of different swelling and, relying on scaling laws. It only uses coarse-grained quantities like the distance between two grafting points and the overall thickness of the brush. The interaction force is dependent on the separation between the CP surface and the hard surface of the microplastic particle. The brush thickness and grafting distance were obtained by the fit. Therefore, no information in the molecular details of the biomolecules in the eco-corona like molecular weight or monomer size is needed. A model like AdG fulfilling this requirement is necessary as the eco-corona is not made up of a single biomolecule with easily characterizable properties. Instead it is a variety of different proteins, polysaccharides, lipids and humic acids with no clear molecular structure.[19, 23, 35]

To a first approximation, the colloidal probe is supposed to have a bare surface, see below for a discussion of this approximation. Consequently, the data was fitted with the asymmetric AdG model proposed by O'Shea[36] for an interaction of a polymer brush with a plain surface. Four assumptions are made to describe the interaction in AdG theory. i) Brushes are end-grafted. This means, each polymer chain is attached to the surface with only one point of the chain, i.e., no loops or more complex structures play a role. ii) Brushes do not interpenetrate, which can be neglected for the asymmetric case because there is only one brush. iii) The density profile of polymer chains within the brush is uniform and zero outside. This assumption translates in a monodisperse distribution of molar masses of the polymers in the brush. iv) No electrostatic forces are present. Only steric forces due to the fluctuations of the polymer are relevant. Assumption ii) is valid at the beginning of the measurements and iv) for the whole experiment. Due to the unknown molecular structure of the eco-corona, we do not expect that assumptions i) and iii) are fulfilled. However, these assumptions might be effectively fulfilled, i.e., deviations from these assumptions may or may not influence the interaction of the probe with the eco-corona.

The asymmetric AdG model considers two mechanisms to describe the interaction of a polymer brush with an inert surface. i) Polymer chains stretch to avoid overlapping and form brushes. ii) Stretched polymers are an entropic spring and store elastic energy. In equilibrium both mechanisms balance each other. When a second surface is approaching and compressing the brush, the polymer concentration

$$\frac{F_{asym}(D)}{R^*} = \frac{4\pi\, k_B T\, L}{35\, s^3} \cdot \left[7 \cdot \left(\frac{D}{L}\right)^{-\frac{5}{4}} + 5 \cdot \left(\frac{D}{L}\right)^{\frac{7}{4}} - 12\right] \text{ for } D < L$$

increases and an osmotic repulsion arises. Contrary to that, entropic elastic energy is released by chain relaxation. With the help of the Derjaguin relation[37] the interaction force of the two spheres was normalized to two planar surfaces.

Here $R^*$ is the effective radius and describes the contact radius between CP and microplastic particle. The parameter $S$ is the (apparent) average grafting distance from one anchor point of a polymer chain to the next, $k_B T$ the thermal energy at lab temperature (293 K), $L$ the (apparent) brush thickness and $D$ the separation between the two surfaces. We denote the grafting distance and brush thickness as apparent parameters, because the assumptions i) and iii) of the AdG model, mentioned above, are not fulfilled. $D$ is the separation given by the AFM, whereby 0 nm separation is defined as the hard, incompressible contact of the two surfaces. $L$ is the separation at the onset of the repulsive force. The only parameters to be fitted are $s$ and L. We fitted all curves seen in Figure 4 in the normalized force range of 0.1 – 4 mN/m. The results are given in Figure 5. The apparent brush thickness $L$ is varying from 15 to 204 nm. These values do not reflect the true thickness of the eco-corona, but give an idea about the dimensions. This will be discussed later on in more detail. The apparent grafting distance $s$ is within the range of 4 to 18 nm. A mean grafting distance of $s$ = 8.9 nm was calculated which corresponds to a grafting density of $1/s^2$ = 0.023 nm$^{-2}$.

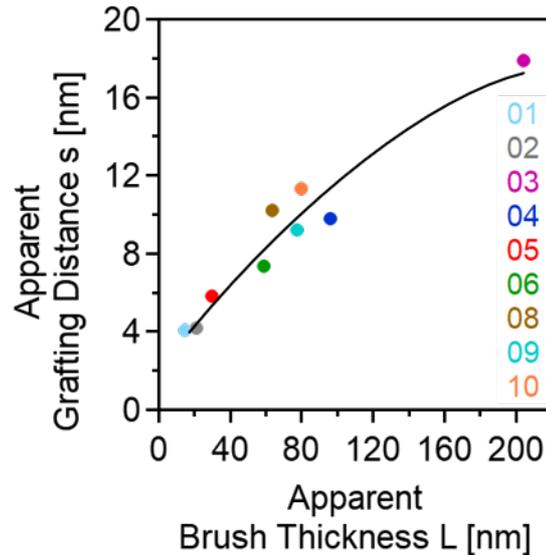

*Figure 5: Increasing grafting distance of the eco-corona with brush thickness. Thin films are more compressed while thick ones are likely more hydrated. Fit is guide to the eye only.*

As depicted in Figure 5, the apparent grafting distance and the apparent brush thickness are not independent from each other, but strongly correlate with each other. Especially, the apparent brush thickness is much higher than the apparent grafting distance. Along the lines of the AdG model, this implies that the assumed brush configuration is actually reflected in the fitted parameters ($s \ll L$). Moreover, there seems to be a small parameter corridor for possible apparent grafting distances in dependence on the apparent brush thickness.

The overall image from these fits in Figure 4 is that the AdG model is able to describe the repulsive interaction of an eco-corona. This observation requires however a further discussion of the assumptions used in the AdG model. Above, we identified two of the assumptions as not obviously fulfilled: i) Brushes are end-grafted. This criterion is obviously not met. The physisorption of the water-soluble biomolecules induces the attachment to the surface with more than one point of the molecule chain. Others might only have (chemical or physical) crosslinks to physisorbed chains. Hence, the conformation of the eco-corona is different to that of a polymer brush. Interestingly, physisorbed block-copolymers[38] show grafting distances and brush thicknesses in the same range as our eco-corona with the thinnest layers. Polyelectrolyte multilayers of poly(styrene sulfonate) (PSS) and poly(diallyldimethylammonium chloride) (PDADMAC) were investigated by Mohamad et al..[39] They obtained a mean grafting distance of 43.2 nm and a high brush thicknesses of more than 100 nm at low ionic strengths. These examples illustrate that the AdG model in fact can give meaningful results for non-brush systems that are indeed in a similar range as our values for the eco-corona. ii) Brushes do not interpenetrate. This assumption could be generally neglected for the asymmetrical AdG model as there is only one surface bearing a polymer brush. In our case however, the asymmetrical model is the better approximation, although we have shown in **Error! Reference source not found.** that biomolecules of the eco-corona attach to the CP during the experiment. If the transferred material would have a similar thickness as the eco-corona, the symmetric system should be more appropriate as the two opposing surfaces are covered. The fits, however, result in a better agreement of the interaction profile with the asymmetric model. From this we conclude that the amount of eco-corona transferred to the CP is not enough to affect the repulsive interaction. iii) Uniform density profile of the polymer chains within the brush. The eco-corona is a natural and hence very heterogeneous system as already discussed. It is likely made up of different biomolecules with different moieties, molecular weight and conformations. Therefore, it is unlikely that the density profile of the eco-corona is uniform and all biomolecule chains protrude equally long into the solution. We attribute the differences between the model and the measured data in Figure 4 at low forces and high separation distances to a decreased polymer concentration in this part of the eco-corona. This non-constant polymer concentration can cause differences between the actual thickness of the eco-corona and the fitted apparent brush thickness. iv) No electrostatic forces are present. The measurements were performed at 150 mM and the electrostatic interactions are therefore screened. Hence, only steric repulsion was observed.

In our study, apparent grafting distances increased with the apparent brush thickness of the eco-corona. Thus, the intermolecular distance between biomolecules increases at high brush thickness. This indicates that the corona is becoming softer the more biomolecules adsorb to the surface. This is in agreement with the proposed model of the "hard" and "soft corona" by Monopoli et al..[40] The term "hard corona" corresponds to the adsorption of biomolecules with high surface affinity. These are supposed to bind strongly onto the surface, which would result in a more rigid shell. The soft corona is forming due to attachment of biomolecules onto the pre-formed hard corona. They are more loosely bound and build therefore a softer second shell around the microplastic particles. However, our data suggest a gradual transition between hard and soft corona.

## 5. Conclusion

Here, we demonstrated with CP-AFM measurements that the formation of the eco-corona on microplastic particles introduces a soft film which changes the mechanical behaviour on the surface. This heterogenous film is different for individual particles but the obtained parameters grafting distance and brush thickness always correlate. Thereby we could mechanically support the gradual transition from the hard to the soft corona. Moreover, we showed that the eco-corona formation around particles lead to the onset of long-range repulsive surface interactions. For the first time, we have shown that the Alexander-de Gennes model can describe the distance-dependency of these interactions well. This theory was originally developed to describe the interaction of polymer brushes but it was successfully applied to adsorbed polymer layers too.[29-32, 38] Although the eco-corona on microplastic particles has a more complex structure, the essential physics of the long-range interactions can be captured by an effective brush description using an apparent brush thickness and apparent grafting distance as model parameters.

Our study contributes to a better understanding of the eco-corona on microplastic particles. The formation of the eco-corona on the microplastic surface changes its properties. We have shown that the higher the thickness of the eco-corona the softer the surface becomes. This correlation is expected to influence the particle-cell interactions and, as Hartmann et al. demonstrated, cellular uptake.[24] The stiffness-dependent uptake of microparticles potentially explains the increased internalisation of eco-corona particles found by Ramsperger et al., too.[23] However, it is not clear whether other surface properties contribute to this effect too. These results make it even more important to shift the focus on realistic microplastic samples covered by an eco-corona. It also highlights the importance to characterise the surface properties of microplastic particles as they influence cellular interactions which may in turn be responsible for adverse effects on environmental and human health.

More importantly we have shown in our study that the physical structure of the eco-corona can be described by a polymer brush model. This fundamental understanding leads to certain implications which can be drawn from their inherent properties. These are swelling behaviour and interaction with other surfaces. Information about these properties is important to assess their transport and bioavailability. Because, it enables to make predictions about the aggregation behaviour of eco-corona covered microplastic particles. Polymer brush interactions could be therefore used in a coarse-grained description for the eco-corona. An effective brush will facilitate modelling its impact with a much lower computational cost than explicitly modelling polymer adsorption layers, opening the door towards larger system simulations.

Interestingly, the effective brush parameters are in a similar range as found for adsorbed polyelectrolyte (multi)layers and polymers.[30, 39, 41] Polyelectrolyte multilayers can be tuned in their mechanical and surface charge properties and are easily applicable to different geometries and sizes.[42, 43] Additionally, the tunability of well-defined polyelectrolytes enables to account effects solely to surface charge or mechanical properties. This suggests that polyelectrolyte multilayers coated microplastics can serve as model systems for studying its interaction with e.g. cells.

## Acknowledgements

The authors acknowledge funding from the Deutsche Forschungsgemeinschaft (DFG; German Research Foundation) project number 391977956–SFB 1357 Microplastics and 422852551 within the priority program 2171.

AFRM.R. and S.W. were supported by the elite network of Bavaria (S.W.: Study Program Biological Physics; AFRM.R.: scholarship of the BayEFG) and the University of Bayreuth Graduate School.

We thank Thomas Scheibel and Hendrik Bargel for the support with the SEM. The SEM was funded by the Deutsche Forschungsgemeinschaft (DFG GZ:INST 91/366-1 FUGG).